\documentclass[11pt]{article}
\usepackage{a4wide,cite,graphicx,latexsym,amssymb}
\usepackage{color}

\newcommand{\rd}{{\rm d}}

\newcommand{\wt}{\widetilde}

\newcommand{\IM}{\mbox{\rm Im}}
\newcommand{\RE}{\mbox{\rm Re}}
\newcommand{\eqn}[1]{(\ref{#1})}
\newcommand{\mev}{\mbox{\rm MeV}}
\newcommand{\gev}{\mbox{\rm GeV}}

\begin{document}


\begin{titlepage}


\begin{flushright}
{UAB-FT-653}\\[16mm]
\end{flushright}

\begin{center}
\begin{boldmath}
{\LARGE\sf\bf $K\pi$ vector form factor, dispersive constraints\\[3mm]
 and $\tau \to \nu_\tau K\pi$ decays}\\[12mm]
\end{boldmath}

{\large\bf Diogo R. Boito${}^{1}$, Rafel Escribano${}^{1}$, and}
{\large\bf Matthias Jamin${}^{1,2}$}\\[10mm]

{\small\sl ${}^{1}$ Grup de F\'{\i}sica Te\`orica and IFAE, Universitat
 Aut\`onoma de Barcelona,\\ E-08193 Bellaterra (Barcelona), Spain.}\\[3mm]
{\small\sl ${}^{2}$ Instituci\'o Catalana de Recerca i Estudis
 Avan\c{c}ats (ICREA).}\\[2cm]
\end{center}

{\bf Abstract:}
Recent experimental data for the differential decay distribution of the decay
$\tau^-\to\nu_\tau K_S\pi^-$ by the Belle collaboration are described by a
theoretical model which is composed of the contributing vector and scalar form
factors $F_+^{K\pi}(s)$ and $F_0^{K\pi}(s)$. Both form factors are constructed
such that they fulfil constraints posed by analyticity and unitarity. A good
description of the experimental measurement is achieved by incorporating two
vector resonances and working with a three-times subtracted dispersion relation
in order to suppress higher-energy contributions. The resonance parameters of
the charged $K^*(892)$ meson, defined as the pole of $F_+^{K\pi}(s)$ in the
complex $s$-plane, can be extracted, with the result
$M_{K^*}=892.0 \pm 0.9\,$MeV and $\Gamma_{K^*}=46.2 \pm 0.4\,$MeV. Finally,
employing a three-subtracted dispersion relation allows to determine the
slope and curvature parameters $\lambda_+^{'}=(24.7\pm 0.8)\cdot 10^{-3}$
and $\lambda_+^{''}=(12.0\pm 0.2)\cdot 10^{-4}$ of the vector form factor
$F_+^{K\pi}(s)$ directly from the data.

\vfill


\end{titlepage}

\newpage
\setcounter{page}{1}


\section{Introduction}

Hadronic decays of the $\tau$ lepton provide a fruitful environment to study
low-energy QCD under rather clean conditions
\cite{bnp92,bra89,bra88,np88,pich89}. Many fundamental QCD parameters can be
determined from investigations of the $\tau$ hadronic width as well as
invariant mass distributions. A prime example in this respect is the QCD
coupling $\alpha_s$ \cite{ddhmz08,bj08,mal08}. In addition, fundamental
parameters of the strange sector in the Standard Model can also be obtained
from the $\tau$ strange spectral function. The experimental separation of the
Cabibbo-allowed decays and Cabibbo-suppressed modes into strange particles
\cite{aleph99,opal04,dhz05} paved the way for a new determination of the
quark-mixing matrix element $|V_{us}|$ \cite{mwbnr08,gjpps07,gjpps04,gjpps03}
as well as the mass of the strange quark
\cite{pp98,ckp98,pp99,kkp00,km00,dchpp01,cdghpp01,bck04}.

Cabibbo-suppressed $\tau$-decays  into strange final states are dominated
by $\tau\to \nu_\tau K\pi$. In the past, ALEPH \cite{aleph99} and OPAL
\cite{opal04} have measured the corresponding distribution function but lately
$B$-factories have become a new source of high-statistics data for this
reaction. Recently, the Belle experiment published results for the
$\tau\to \nu_\tau K\pi$ spectrum \cite{belle07} and a new determination of
the total branching fraction became available from BaBar
\cite{babar07,babar08,wren08}. In the future, there are good prospects for
results on the full spectrum both from BaBar and BESIII.

Theoretically, the general expression for the differential decay distribution
of the decay $\tau\to\nu_\tau K\pi$ can be written as \cite{fm96}
\begin{equation}
\label{dGtau2kpi}
\frac{\rd\Gamma_{K\pi}}{\rd\sqrt{s}} \,=\, \frac{G_F^2|V_{us}|^2 M_\tau^3}
{32\pi^3s}\,S_{\rm EW}\biggl(1-\frac{s}{M_\tau^2}\biggr)^{\!2}\Biggl[
\biggl( 1+2\,\frac{s}{M_\tau^2}\biggr) q_{K\pi}^3\,|F_+^{K\pi}(s)|^2 +
\frac{3\Delta_{K\pi}^2}{4s}\,q_{K\pi}|F_0^{K\pi}(s)|^2 \Biggr] ,
\end{equation}
where isospin invariance is assumed   and we have summed over the two possible
decay channels $\tau^-\to\nu_\tau\overline K^0\pi^-$ and
$\tau^-\to\nu_\tau K^-\pi^0$, with the individual decays contributing in the
ratio $2\hspace{-0.4mm}:\!1$ respectively. Furthermore, $S_{\rm EW}=1.0201$
\cite{erl02} represents an electro-weak correction factor,
$\Delta_{K\pi}\equiv M_K^2-M_\pi^2$, and $q_{K\pi}$ is the kaon momentum in
the rest frame of the hadronic system,
\begin{equation}
q_{K\pi}(s) \,=\, \frac{1}{2\sqrt{s}}\sqrt{\Big(s-(M_K+M_\pi)^2\Big)
\Big(s-(M_K-M_\pi)^2\Big)}\cdot\theta\Big(s-(M_K+M_\pi)^2\Big) \,.
\end{equation}
Finally, we denote by $F_+^{K\pi}(s)$ and $F_0^{K\pi}(s)$ the vector and scalar $K\pi$ form factors respectively, which we will discuss in detail below. 

In eq.~\eqn{dGtau2kpi}, the prevailing contribution is due to the $K\pi$ vector
form factor $F_+^{K\pi}(s)$, and in the energy region of interest, this form
factor is by far dominated by the $K^*(892)$ meson. A description of
$F_+^{K\pi}(s)$ based on the chiral theory with resonances (R$\chi$T)
\cite{egpr89,eglpr89} was provided in ref.~\cite{jpp06}, analogous to a similar
description of the pion form factor presented in refs.~\cite{gp97,pp01,scp02}.
Then in ref.~\cite{jpp08} this description was employed in fitting Belle data
for the spectrum of the decay $\tau\to\nu_\tau K_S\pi^-$ \cite{belle07}.
The additionally required scalar $K\pi$ form factor $F_0^{K\pi}(s)$ had been
calculated in the same R$\chi$T plus dispersive constraint framework in a
series of articles \cite{jop00,jop01a,jop01b}, and the recent update of
$F_0^{K\pi}(s)$ \cite{jop06} was incorporated as well.

A slight drawback of the description for the vector form factors of
refs.~\cite{gp97,jpp06} is that the form factors only satisfy the analyticity
constraints in a perturbative sense, that is up to higher orders in the chiral
expansion. Though the violation of analyticity is expected to only be a small
correction (of order $p^6$ in the chiral expansion in the case at hand) it is
certainly worthwhile to corroborate this assumption. A coupled channel analysis
of the $K\pi$ vector form factor, which would allow for such a test, was
performed in ref.~\cite{mou07}. However, in ref.~\cite{mou07} the theoretical
description was not really fitted to the experimental data, so that it is
difficult to decide if differences of the coupled channel analysis \cite{mou07}
as compared to the description \cite{jpp06,jpp08} already show up in the
current experimental data. Even more so as the fits performed in
ref.~\cite{jpp08} provided a satisfactory description of the Belle spectrum.

Below, we shall investigate the related questions in a more modest approach.
In the region of the $K^*(892)$ meson, elastic unitarity is still expected to
hold. Since this meson dominates the $K\pi$ vector form factor, an ansatz
implementing elastic unitarity should result in a good approximation. For the
pion vector form factor such an approach was pursued in ref.~\cite{pp01} and
in the present work we perform an analogous investigation for $F_+^{K\pi}(s)$.
Even though possible coupled-channel contributions are not explicitly included
in our parametrisation of the $K\pi$ vector form factor, their influence can
be studied through the sensitivity of our ansatz when changing the number of
subtractions in the dispersion relation that the form factor satisfies, because
a larger number of subtractions entails a stronger suppression of higher-energy
contributions. As a benefit of our approach, we are able to extract the
resonance parameters of the $K^*(892)$ from the pole of $F_+^{K\pi}(s)$ in the
complex $s$-plane, which should be regarded as more model-independent than
the Breit-Wigner type parameters extracted in the previous analyses
\cite{belle07,jpp08}, as well as the first three slopes in the Taylor
expansion of $F_+^{K\pi}(s)$ around $s=0$.

\begin{boldmath}
\section{The $K\pi$ vector form factor}
\end{boldmath}

The $K\pi$ vector form factor $F_+^{K\pi}(s)$ is an analytic function in the
complex $s$-plane, except for a cut along the positive real axis, starting at
the $K\pi$ threshold $s_{K\pi}\equiv (M_K+M_\pi)^2$, where its imaginary part
develops a discontinuity. The analyticity and unitarity properties of the form
factor result in the fact that it satisfies an $n$-subtracted dispersion
relation, explicated in more detail for example in refs.~\cite{gp97,pp01}. In
the elastic region below roughly $1.2\;\gev$, the dispersion relation admits
the well-known Omn\`es solution \cite{om58}
\begin{equation}
\label{omel}
F_+^{K\pi}(s) \,=\, P_n(s) \exp \Biggl\{ \frac{s^n}{\pi}\!
\int\limits^\infty_{s_{K\pi}}\!\!ds'\, \frac{\delta_1^{K\pi}(s')}
{(s')^n(s'-s-i0)}\Biggr\} \,,
\end{equation}
which corresponds to performing the $n$ subtractions at $s=0$, and where
\begin{equation}
\label{Pns}
P_n(s) \,=\, \exp\Biggl\{\,\sum\limits_{k=0}^{n-1} \,\frac{s^k}{k!}\,
\frac{d^k}{ds^k} \ln F_+^{K\pi}(s)\biggr|_{s=0} \,\Biggr\}
\end{equation}
is the subtraction polynomial. More general formulae with subtractions at an
arbitrary point $s=s_0$ can for example be found in ref.~\cite{pp00}.
Furthermore, $\delta_1^{K\pi}(s)$ is the P-wave $I=1/2$ elastic $K\pi$ phase
shift. As on general grounds the phase shift is expected to go to an integer
multiple of $\pi$, at least one subtraction ($n=1$) is required in
eq.~\eqn{omel} to make the integral convergent. Let us first dwell on this
case in more detail, before turning to the case with a larger number of
subtractions.

While in the approach of refs.~\cite{gp97,jpp06} to the vector form factor
the real part of the one-loop integral function $\wt H(s)$ (for its definition
see \cite{jpp06}) is resummed into an exponential, strict unitarity is only
maintained if this piece, together with the imaginary part which provides
the width of the resonance, is resummed in the denominator of the form factor
\cite{oor00,jop00}. The resulting expression of the vector form factor
corresponding to one single resonance then reads
\begin{equation}
\label{Fpone}
F_+^{K\pi}(s) \,=\, \frac{m_{K^*}^2}{m_{K^*}^2 - s -
\kappa \,\wt H_{K\pi}(s)} \,.
\end{equation}
Of course, up to order $p^4$ in the chiral expansion, resumming the real part
in the denominator or an exponential is fully equivalent. Differences of both
approaches first start to appear at ${\cal O}(p^6)$. In eq.~\eqn{Fpone} the
parameter $m_{K^*}$ is to be distinguished from the true mass of the $K^*$
meson $M_{K^*}$, which later will be identified with the real part of the
pole position of $F_+^{K\pi}(s)$ in the complex $s$-plane.\footnote{The
renormalisation scale $\mu$ appearing in $\wt H_{K\pi}(s)$ will be set to
the physical resonance mass $\mu=M_{K^*}$.}

Identifying the imaginary part in the denominator of eq.~\eqn{Fpone} with
$-\,m_{K^*}\gamma_{K^*}(s)$, where the $s$-dependent width of the $K^*$ meson
takes the generic form of a vector resonance,
\begin{equation}
\label{GaKs}
\gamma_{K^*}(s) \,=\, \gamma_{K^*}\frac{s}{m_{K^*}^2} \frac{\sigma^3_{K\pi}(s)}
{\sigma^3_{K\pi}(m_{K^*}^2)} \,,
\end{equation}
with $\gamma_{K^*}\equiv \gamma_{K^*}(m_{K^*}^2)$, the dimensionful constant
$\kappa$ has to take the value:
\begin{equation}
\label{kappa}
\kappa \,=\, \frac{192\pi F_K F_\pi}{\sigma_{K\pi}(m_{K^*}^2)^3}
\frac{\gamma_{K^*}}{m_{K^*}} \,.
\end{equation}
In eqs.~\eqn{GaKs} and \eqn{kappa}, the phase space function $\sigma_{K\pi}(s)$
is given by $\sigma_{K\pi}(s)=2q_{K\pi}(s)/\sqrt{s}$. The form factor
$F_+^{K\pi}(s)$ can therefore be written in the equivalent form
\begin{equation}
\label{Fponea}
F_+^{K\pi}(s) \,=\, \frac{m_{K^*}^2}{m_{K^*}^2 - s -
\kappa \,\RE\wt H_{K\pi}(s) - i\, m_{K^*}\gamma_{K^*}(s)} \,.
\end{equation}
From eq.~\eqn{Fponea} the normalisation $F_+^{K\pi}(0)$ at $s=0$ which is
needed in order to calculate the reduced form factor
$\wt F_+^{K\pi}(s)\equiv F_+^{K\pi}(s)/F_+^{K\pi}(0)$, is given by
\begin{equation}
\label{Fp0}
F_+^{K\pi}(0) \,=\, \frac{m_{K^*}^2}{m_{K^*}^2 - \kappa\,\wt H_{K\pi}(0)} \,.
\end{equation}
The phase $\delta_1^{K\pi}(s)$ of the form factor $F_+^{K\pi}(s)$ is found
to be:
\begin{equation}
\label{delta1KP}
\tan\delta_1^{K\pi}(s) \,\equiv\, \frac{\IM F_+^{K\pi}(s)}{\RE F_+^{K\pi}(s)}
\,=\, \frac{m_{K^*}\gamma_{K^*}(s)}{m_{K^*}^2-s -
\kappa \,\RE\tilde H_{K\pi}(s)} \,.
\end{equation}
It is a straightforward matter to verify that with the phase
$\delta_1^{K\pi}(s)$ of eq.~\eqn{delta1KP},  the form factor of
eq.~\eqn{Fponea} satisfies the Omn\`es relation \eqn{omel} with $n=1$ and
$P_1(s) = F_+^{K\pi}(0)$. Therefore, it is in accord with the analyticity
and unitarity requirements. Finally, the reduced form factor is written as
\begin{equation}
\label{Fptilde}
\wt F_+^{K\pi}(s)\,=\, \frac{m_{K^*}^2 - \kappa\,\wt H_{K\pi}(0)}{m_{K^*}^2 -
s - \kappa \,\RE\wt H_{K\pi}(s) - i\, m_{K^*}\gamma_{K^*}(s)} \,,
\end{equation}
which resembles the non-strange form factor of Gounaris and Sakurai
\cite{Gounaris:1968mw}. (For comparison, see also ref.~\cite{Kuhn:1990ad}.)
However, while in the Breit-Wigner resonance shape of
ref.~\cite{Gounaris:1968mw} the real part of the inverse propagator is
obtained through a twice subtracted dispersion relation combined with proper
subtractions to fix mass and width of the resonance, the form factor in
eq.~\eqn{Fptilde} is found from a once subtracted dispersion relation
satisfying analyticity and unitarity.

As has already been discussed in ref.~\cite{jpp08}, in eq.~\eqn{dGtau2kpi}
only the reduced form factor $\wt F_+^{K\pi}(s)$ has to be modelled, as the
normalisation of $F_+^{K\pi}(s)$ only appears in the product
$|V_{us}| F_+^{K\pi}(0)$. This combination is determined most precisely from
the analysis of semi-leptonic kaon decays. A  recent average was presented by
the FLAVIAnet kaon working group, and reads \cite{FKWG08}
\begin{equation}
\label{VusF0}
|V_{us}| F_+^{K^0\pi^-}(0) \,=\, 0.21664 \pm 0.00048 \,.
\end{equation}
In what follows, we work  with form factors normalised to one at the origin
and assume the value \eqn{VusF0} for the overall normalisation. We remark that
the normalisation for the scalar and the vector form factors is the same and
that \eqn{VusF0} already corresponds to the $K^0\pi^-$ channel, which was
analysed by the Belle collaboration \cite{belle07}. Consequently, possible
isospin-breaking corrections to $F^{K^0\pi^-}_+(0)$ are properly taken into
account.

\section{Single resonance fits to the Belle spectrum}

Our fits to the Belle $\tau^-\to\nu_\tau K_S\pi^-$ spectrum \cite{belle07} will
be performed in complete analogy to the recent analysis of ref.~\cite{jpp08}.
Let us briefly review the main strategy for these fits. The central fit
function is taken to have the form
\begin{equation}
\label{FitFun}
\frac{1}{2}\cdot \frac{2}{3}\cdot 0.0115\,[{\rm GeV/bin}]\,{\cal N}_T\cdot
\frac{1}{\Gamma_\tau \bar B_{K\pi}}\,\frac{\rd\Gamma_{K\pi}}{\rd\sqrt{s}} \,.
\end{equation}
The factors $1/2$ and $2/3$ arise because the $K_S\pi^-$ channel has been
analysed. Then, $11.5\,$MeV was the bin-width chosen by the Belle collaboration,
and ${\cal N}_T=53110$ is the total number of observed signal events. Finally,
$\Gamma_\tau$ denotes the total decay width of the $\tau$ lepton and
$\bar B_{K\pi}$ a remaining normalisation factor that will be deduced from the
fits. The normalisation of our ansatz \eqn{FitFun} is taken such that for a
perfect agreement between data and fit function, $\bar B_{K\pi}$ would
correspond to the total branching fraction
$B_{K\pi}\equiv B[\tau^-\to\nu_\tau K_S\pi^-]$ which is obtained by integrating
the decay spectrum. All further numerical input parameters have been chosen as
in ref.~\cite{jpp08}.

\begin{table}[htb]
\begin{center}
\vspace{3mm}
\begin{tabular}{c c}
\hline
\hline
&  Eq.~\eqn{Fponea} for $F_+^{K\pi}(s)$\\
\hline
$\bar B_{K\pi}$ ($B_{K\pi}$) &  $0.3611 \pm 0.0042\,\% \,(0.3562\,\%)$ \\
$m_{K^*}$ & $943.35\pm 0.51 \,\mbox{MeV}$ \\
$\gamma_{K^*}$ & $66.29 \pm 0.79 \,\mbox{MeV}$ \\
\hline
$\chi^2/{\rm n.d.f.}$ & 39.4/27\\
\hline
\hline
\end{tabular}
\caption{ Results for the fit with a one-subtracted dispersion relation
including a single vector resonance in $F_+^{K\pi}(s)$ according to
eq.~\eqn{Fponea}, as well as the scalar form factor $F_0^{K\pi}(s)$
\cite{jop06}.\label{tab1}}
\end{center}
\end{table}

Numerically, our first fit of the Belle data \cite{belle07} with the vector
form factor $F_+^{K\pi}(s)$ according to eq.~\eqn{Fponea}, and including data
up to $\sqrt{s}=1.01525\;\gev$ (centre of bin 34), is presented in
table~\ref{tab1}. As discussed in ref.~\cite{jpp08}, we have removed the
problematic data points 5, 6 and 7 from the fit. Furthermore, we have not
taken into account the lowest data point since for our physical meson masses
the centre of this bin lies below the $K\pi$ threshold. For the scalar form
factor $F_0^{K\pi}(s)$ we have used the recent update \cite{jop06}, employing
the central parameters given there. One observes that due to the not very
satisfactory quality of the fit, the fit parameter $\bar B_{K\pi}$ and the
integrated branching fraction $B_{K\pi}$ display a marked deviation, which is
however within the uncertainties.  Furthermore, because of the real part of
the loop-integral in the denominator, $m_{K^*}$ and $\gamma_{K^*}$ turn out
rather different from their physical values $M_{K^*}$ and $\Gamma_{K^*}$. For
the final results of the parameters in our description of the vector form
factor, in our conclusion we shall also present values for the physical
parameters $M_{K^*}$ and $\Gamma_{K^*}$, as obtained from the pole position
in the complex $s$-plane.

In order to make the fit less sensitive to deficiencies of our description in
the higher-energy region, a larger number of subtractions can be applied to the
dispersion relation. Besides, employing an $n$-subtracted dispersion relation
has the advantage that the slope parameters which appear as subtraction
constants are determined more directly from the data. It should be pointed out,
however, that the form factor with a larger number of subtractions $n\geq 2$
violates the expected QCD large-energy behaviour. For large enough energies
the form factor $F_+^{K\pi}(s)$ should vanish as $1/s$. Since we only employ
the vector form factor up to about $\sqrt{s}\approx 1.7\,\gev$, which is still
in the resonance region, we consider this deficiency acceptable. Anyhow, as
can be verified explicitly from our fits, in the considered region above the
second vector resonance, $F_+^{K\pi}(s)$ is a decreasing function of $s$. On
the other hand, fits with only one subtraction were generally found to only
provide a poor description of the experimental data.

Let us present our results for the case $n=3$ in detail, but below, we shall
also briefly comment on the cases $n=2$ and $n=4$. The importance of the
high-energy region can be studied by introducing a cutoff $s_{\rm cut}$ as
the upper limit of the integration in the Omn\`es integral \eqn{omel}.
Incorporating three subtractions, the reduced form factor $\wt F_+^{K\pi}(s)$
then takes the form:
\begin{equation}
\label{Ftil3sub1res}
\wt F_+^{K\pi}(s) \,=\, \exp\Biggr\{ \alpha_1 \frac{s}{M_{\pi^-}^2} +
\frac{1}{2}\alpha_2\frac{s^2}{M_{\pi^-}^4} + \frac{s^3}{\pi}\!
\int\limits^{s_{\rm cut}}_{s_{K\pi}} \!\!ds'\, \frac{\delta_1^{K\pi}(s')}
{(s')^3(s'-s-i0) }\Biggr\} \,,
\end{equation}
where the phase $\delta_1^{K\pi}(s)$ corresponds to the expression of
eq.~\eqn{delta1KP}. The parameters $\alpha_1$ and $\alpha_2$ can easily be
related to the slope parameters $\lambda_+^{(n)}$, which appear in the Taylor
expansion of $\wt F_+^{K\pi}(s)$ around $s=0\,$:
\begin{equation}
\wt F_+^{K\pi}(s) \,=\, 1 + \lambda_+^{'}\frac{s}{M_{\pi^-}^2} +
\frac{1}{2}\, \lambda_+^{''}\frac{s^2}{M_{\pi^-}^4} +
\frac{1}{6}\,\lambda_+^{'''} \frac{s^3}{M_{\pi^-}^6} + \ldots \,.
\end{equation}
Explicitly, the relations for the linear and quadratic slope parameters
$\lambda_+^{'}$ and $\lambda_+^{''}$ then take the form:
\begin{equation}
\label{lambdaalpha}
\lambda_+^{'}  \,=\, \alpha_1 \,, \qquad
\lambda_+^{''} \,=\, \alpha_2 + \alpha_1^2 \,.
\end{equation}
Below, we shall also compute the cubic slope parameter $\lambda_+^{'''}$ from
the dispersive integral.

\begin{table}[htb]
\begin{center}
\vspace{3mm}
\begin{tabular}{c c c c c}
\hline
\hline
& $s_{\rm cut} = 3.24\;\gev^2$ & $s_{\rm cut} = 4\;\gev^2$ &
  $s_{\rm cut} = 9\;\gev^2$ & $s_{\rm cut} \to \infty$ \\
\hline
$\bar B_{K\pi}$ &  $0.394 \pm 0.045\,\%$ & $0.397 \pm 0.046\,\%$ &
                   $0.398 \pm 0.046\,\%$ & $0.398 \pm 0.046\,\%$ \\
$(B_{K\pi})$ & $(0.389\,\%)$ & $(0.391\,\%)$ & $(0.393\,\%)$ & $(0.393\,\%)$ \\
$m_{K^*}\;[\mev]$ & $943.34 \pm 0.57$ & $943.36 \pm 0.58$ &
                    $943.37 \pm 0.58$ & $943.37 \pm 0.58$ \\
$\gamma_{K^*}\;[\mev]$ & $66.48 \pm 0.87$ & $66.50 \pm 0.89$ &
                         $66.52 \pm 0.89$ & $66.52 \pm 0.89$ \\
$\lambda_+^{'} \times 10^{3}$ & $23.9 \pm 2.4$ & $24.2 \pm 2.5$ &
                                $24.4 \pm 2.5$ & $24.4 \pm 2.5$ \\
$\lambda_+^{''}\times 10^{4}$ & $11.5 \pm 0.7$ & $11.5 \pm 0.7$ &
                                $11.5 \pm 0.7$ & $11.5 \pm 0.7$ \\
\hline
$\chi^2/{\rm n.d.f.}$ & 45.8/41 & 45.8/41 & 45.7/41 & 45.8/41 \\
\hline
\hline
\end{tabular}
\caption{ Results for the fits with a three-subtracted dispersion relation
including a single vector resonance in $F_+^{K\pi}(s)$ according to
eq.~\eqn{Fponea}, as well as the scalar form factor $F_0^{K\pi}(s)$
\cite{jop06}.\label{tab2}}
\end{center}
\end{table}

The results of our fits with the three-subtracted dispersion relation,
employing four values of $\sqrt{s_{\rm cut}}$, namely $1.8\;\gev$, $2\;\gev$,
$3\;\gev$ and $s_{\rm cut} \to \infty$, are given in table~\ref{tab2}. For
these fits, we have included experimental data up to the data point $50$ at
$\sqrt{s}=1.19925\;\gev$ (centre of the bin) \cite{belle07}, and as already
mentioned above, removing the problematic data points 5, 6 and 7. As can be
seen from table~\ref{tab2}, the fits are indeed rather insensitive to the
upper integration limit $s_{\rm cut}$, implying that the higher-energy region
is well suppressed.  Compared with the fit of  table~\ref{tab1}, which only
employed a single subtraction, with $\chi^2/{\rm n.d.f.}\approx 1.1$ the fit
quality is substantially improved. Related to that, also the results for
$\bar B_{K\pi}$ and $B_{K\pi}$ turn out much closer. However, the slope
parameter $\lambda_+^{'}$ is not well determined from our fit. This originates
in the fact that the slope parameters are almost 100\,\% correlated with the
total branching fraction $\bar B_{K\pi}$, and this parameter has relatively
large uncertainties. Therefore, in our best estimate of the model parameters
below, we shall impose the experimental measurement of the total branching
fraction $B_{K\pi}$.

Though we only present explicit results for the three-subtracted dispersion
relation, we have also investigated the cases $n=2$ and $n=4$. In the case
$n=2$, a still somewhat stronger dependence on the cutoff $s_{\rm cut}$ is
observed, which is why we do not discuss the corresponding results in detail.
The four-subtracted dispersion relation, on the other hand, yields almost
unchanged central values for the fit parameters without any improvement in the
$\chi^2/{\rm n.d.f.}$, but with larger parameter errors due to the additional
degree of freedom. From this we conclude that the case $n=3$ discussed above
is an optimal choice as far as the number of subtractions is concerned. Next,
we shall also include a second vector resonance, the $K^*(1410)$ into our
description of the $K\pi$ vector form factor.

\section{Fits with two vector resonances}

For our final fits, we aim at a description of the $\tau\to\nu_\tau K\pi$
spectrum in the full energy range up to the $\tau$ mass. To this end, we also
introduce a second vector resonance, the $K^{*'}=K^*(1410)$, into our model
for the vector form factor $F_+^{K\pi}(s)$. As it is unclear how to directly
incorporate a second vector resonance in the phase $\delta_1^{K\pi}(s)$, we
have followed a somewhat indirect approach, which should be sufficient for our
purposes, as the contribution of the $K^{*'}$ resonance is suppressed by phase
space.

Like for the case of the single resonance, as a starting point we assume the
form of $F_+^{K\pi}(s)$ given in eq.~(5) of ref.~\cite{jpp08}, however resumming
the real part of $\wt H_{K\pi}(s)$ in the denominator of the form factor.
This leads to the following expression:
\begin{equation}
\label{FpKpi2}
\wt F_+^{K\pi}(s) \,=\, \frac{m_{K^*}^2 - \kappa_{K^*}\,\wt H_{K\pi}(0) +
\gamma s}{D(m_{K^*},\gamma_{K^*})} -
\frac{\gamma s}{D(m_{K^{*'}},\gamma_{K^{*'}})} \,,
\end{equation}
where
\begin{equation}
\label{Dden}
D(m_n,\gamma_n) \,\equiv\, m_n^2 - s - \kappa_n \RE\wt H_{K\pi}(s) -
i\, m_n \gamma_n(s) \,.
\end{equation}
For both resonances, $\gamma_n(s)$ is given equivalently to the form of
eq.~\eqn{GaKs}, and the corresponding $\kappa_n$ can be deduced in analogy to
eq.~\eqn{kappa}.  Like in eq.~\eqn{delta1KP}, the phase $\delta_1^{K\pi}(s)$
can be calculated from the relation
\begin{equation}
\label{del1Kpi2}
\tan\delta_1^{K\pi}(s) = \frac{\IM F_+^{K\pi}(s)}{\RE F_+^{K\pi}(s)} \,.
\end{equation}
This is the phase that we then employ in the Omn\`es integral representation
\eqn{omel} for the form factor to perform our fits.

\begin{table}[htb]
\begin{center}
\vspace{3mm}
\begin{tabular}{c c c c c}
\hline
\hline
& $s_{\rm cut} = 3.24\;\gev^2$ & $s_{\rm cut} = 4\;\gev^2$ &
  $s_{\rm cut} = 9\;\gev^2$ & $s_{\rm cut} \to \infty$ \\
\hline
$\bar B_{K\pi}$ &  $0.386 \pm 0.043\%$ & $0.404 \pm 0.044\%$ &
                   $0.417 \pm 0.046\%$ & $0.417 \pm 0.046\%$ \\
$(B_{K\pi})$ & $(0.384\%)$ & $(0.402\%)$ & $(0.414\%)$ & $(0.414\%)$ \\
$m_{K^*}\;[\mev]$ & $943.27 \pm 0.58$ & $943.40 \pm 0.57$ &
                    $943.48 \pm 0.57$ & $943.49 \pm 0.57$ \\
$\gamma_{K^*}\;[\mev]$ & $66.43 \pm 0.90$ & $66.66 \pm 0.87$ &
                         $66.81 \pm 0.86$ & $66.81 \pm 0.87$ \\
$m_{K^{*'}}\;[\mev]$ & $1392 \pm 41$ & $1369 \pm 30$ &
                       $1361 \pm 28$ & $1361 \pm 28$ \\
$\gamma_{K^{*'}}\;[\mev]$ & $273 \pm 137$ & $224 \pm 101$ &
                            $212 \pm  93$ & $212 \pm  93$ \\
$\gamma \times 10^2$ & $-\,4.2 \pm 2.0$ & $-\,3.6 \pm 1.6$ &
                       $-\,3.4 \pm 1.5$ & $-\,3.4 \pm 1.5$ \\
$\lambda_+^{'} \times 10^{3}$ & $22.6 \pm 2.2$ & $23.9 \pm 2.1$ &
                                $24.7 \pm 2.1$ & $24.8 \pm 2.1$ \\
$\lambda_+^{''}\times 10^{4}$ & $11.5 \pm 0.6$ & $11.7 \pm 0.7$ &
                                $11.9 \pm 0.7$ & $11.9 \pm 0.7$ \\
\hline
$\chi^2/{\rm n.d.f.}$ & 73.7/78 & 75.6/78 & 77.2/78 & 77.3/78 \\
\hline
\hline
\end{tabular}
\caption{Results for the fits with a three-subtracted dispersion relation
including two vector resonances in $F_+^{K\pi}(s)$ according to
eqs.~\eqn{FpKpi2} to \eqn{del1Kpi2}, as well as the scalar form factor
$F_0^{K\pi}(s)$ of ref.~\cite{jop06}.\label{tab3}}
\end{center}
\end{table}

Our first fit with two vector resonances proceeds in complete analogy to the
second fit in the single-resonance case. Again, we use the three-subtracted
dispersion relation, and investigate four values of $s_{\rm cut}$, in order to
study the dependence of our fits on the higher-energy contributions. Now, the
Belle data \cite{belle07} have been included up to $\sqrt{s}=1.65925\;\gev$
(data point 90).\footnote{Although the full Belle data set consists of 100 data
points, we follow a suggestion of the experimentalists to only fit the data
up to point 90 \cite{epi07}.} The corresponding fit results are presented in
table~\ref{tab3}.

The general picture of our fits with two resonances is quite satisfying. The
theoretical model provides a good description of the experimental data in the
full energy range. The $\chi^2/{\rm n.d.f.}$ for all values of $s_{\rm cut}$
turns out smaller than one. The dependence of the resulting fit parameters on
$s_{\rm cut}$ is small and within the uncertainties, though clearly visible. 
The fits provide a precise determination of the parameters of the lowest lying
$K^*$ vector resonance, and a still reasonable accuracy for the second
$K^{*'}$ resonance. However, the fit uncertainties for the branching fraction
$\bar B_{K\pi}$ are relatively large. Accordingly, also the error on the slope
parameter $\lambda_+^{'}$ is found large, because these two parameters are
almost $100\,\%$ correlated.

\begin{table}[htb]
\begin{center}
\vspace{3mm}
\begin{tabular}{c c c c c}
\hline
\hline
& $s_{\rm cut} = 3.24\;\gev^2$ & $s_{\rm cut} = 4\;\gev^2$ &
  $s_{\rm cut} = 9\;\gev^2$ & $s_{\rm cut} \to \infty$ \\
\hline
$m_{K^*}\;[\mev]$ & $943.32 \pm 0.59$ & $943.41 \pm 0.58$ &
                    $943.48 \pm 0.57$ & $943.49 \pm 0.57$ \\
$\gamma_{K^*}\;[\mev]$ & $66.61 \pm 0.88$ & $66.72 \pm 0.86$ &
                         $66.82 \pm 0.85$ & $66.82 \pm 0.85$ \\
$m_{K^{*'}}\;[\mev]$ & $1407 \pm 44$ & $1374 \pm 30$ &
                       $1362 \pm 26$ & $1362 \pm 26$ \\
$\gamma_{K^{*'}}\;[\mev]$ & $325 \pm 149$ & $240 \pm 100$ &
                            $216 \pm  86$ & $215 \pm 86$ \\
$\gamma \times 10^2$ & $-\,5.2 \pm 2.0$ & $-\,3.9 \pm 1.5$ &
                       $-\,3.5 \pm 1.3$ & $-\,3.5 \pm 1.3$ \\
$\lambda_+^{'} \times 10^{3}$ & $24.31 \pm 0.74$ & $24.66 \pm 0.69$ &
                                $24.94 \pm 0.68$ & $24.96 \pm 0.67$ \\
$\lambda_+^{''}\times 10^{4}$ & $12.04 \pm 0.20$ & $11.99 \pm 0.19$ &
                                $11.96 \pm 0.19$ & $11.96 \pm 0.19$ \\
\hline
$\chi^2/{\rm n.d.f.}$ & 74.2/79 & 75.7/79 & 77.2/79 & 77.3/79 \\
\hline
\hline
\end{tabular}
\caption{Results for the fits with a three-subtracted dispersion relation
including two vector resonances in $F_+^{K\pi}(s)$ according to
eqs.~\eqn{FpKpi2} to \eqn{del1Kpi2}, as well as the scalar form factor
$F_0^{K\pi}(s)$ \cite{jop06}. The total branching fraction $B_{K\pi}$ has been
fixed to the experimental value \eqn{BKpi}, and the corresponding uncertainty
is included in quadrature.\label{tab4}}
\end{center}
\end{table}

In order to provide a better determination of the slope parameters, a possible
way to proceed is to fix the total branching fraction $B_{K\pi}$ to the
experimental measurement. A very recent update of the world average has been
presented in ref.~\cite{wren08,belle07,babar08}, with the finding:
\begin{equation}
\label{BKpi}
B[\tau^-\to\nu_\tau K_S\pi^-] \,=\, 0.418 \pm 0.011\,\% \,.
\end{equation}
Thus, for our final fits, we have fixed $B_{K\pi}$ to take this value. The
resulting fit parameters for the remaining quantities are displayed in
table~\ref{tab4}. Here, the quoted uncertainties include the statistical fit
errors as well as a variation of $B_{K\pi}$ in the range given by \eqn{BKpi}.
From table~\ref{tab4} we infer that due to the reduction in the uncertainty
of $B_{K\pi}$, correspondingly also the uncertainty in the slope parameters
is much reduced, while the errors of the remaining parameters to a good
approximation stay as before. Also the $\chi^2/{\rm n.d.f.}$ is practically
unchanged, still remaining below one for all values of $s_{\rm cut}$.

\begin{figure}[thb]
\begin{center}
\includegraphics[angle=0, width=15cm]{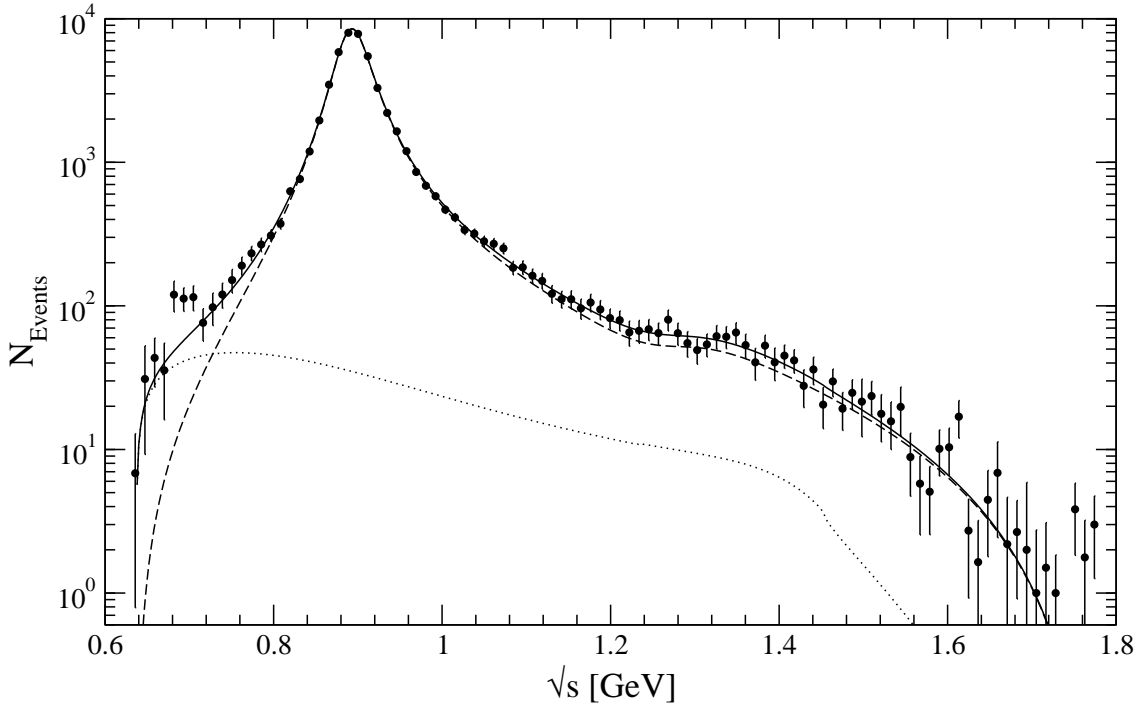}
\end{center}
\caption{Main fit result to the Belle data \cite{belle07} for the differential
decay distribution of the decay $\tau^-\to\nu_\tau K_S\pi^-$. Our theoretical
description corresponds to the fit of table~\ref{tab4} with
$s_{\rm cut} = 4\;\gev^2$. The full fit including vector form factor
$F_+^{K\pi}(s)$ and scalar form factor $F_0^{K\pi}(s)$ is displayed as the
solid line. The separate vector and scalar contributions are shown as the
dashed and dotted lines respectively.\label{fig1}}
\end{figure}

A graphical account of our central result is displayed in figure~\ref{fig1}.
The solid line corresponds to the fit of table~\ref{tab4} including vector
form factor $F_+^{K\pi}(s)$ as well as the scalar form factor $F_0^{K\pi}(s)$
at $s_{\rm cut} = 4\;\gev^2$. The separate contributions of $F_+^{K\pi}(s)$
and $F_0^{K\pi}(s)$ are shown as the dashed and dotted lines respectively.
As is apparent, apart from the data points 5, 6 and 7 in the low-energy region,
our model provides a perfect description of the experimental data by the
Belle collaboration. Also, from an inspection of the region below the $K^*$
resonance it is evident that a contribution from the scalar form factor
$F_0^{K\pi}(s)$ is required, though, like in the analysis of ref.~\cite{jpp08},
the sensitivity to $F_0^{K\pi}(s)$ is not strong enough to allow for a
determination of the corresponding model parameters.

\section{Conclusions}

Hadronic $\tau$ decays provide a means to obtain information on low-energy
QCD as well as hadron phenomenology. In this work, we have studied recent data
on the decay channel $\tau^-\to\nu_\tau K_S\pi^-$ by the Belle collaboration
\cite{belle07}. The measured decay spectrum allows to test models for the
vector and scalar $K\pi$ form factors $F_+^{K\pi}(s)$ and $F_0^{K\pi}(s)$,
and to deduce the corresponding model parameters for the vector form factor.
For $F_+^{K\pi}(s)$, we have used a model which incorporates the constraints
on the form factor from analyticity and elastic unitarity. Furthermore, we
investigated $n$-subtracted dispersive integrals with a cutoff $s_{\rm cut}$,
where $n$ ranges from $1$ to $4$ and $\sqrt{s_{\rm cut}}$ was varied between
$1.8\;\gev$ and infinity. This allowed to test the sensitivity (or
insensitivity) of the form factor model to contributions from higher energies
which are not well known. For the scalar form factor $F_0^{K\pi}(s)$, we have
employed the description of ref.~\cite{jop01a}, which is based on solving
dispersion relations for a two-body coupled-channel problem, and was recently
updated in \cite{jop06}.

Let us begin with summarising our final results for the parameters of the
$K^*$ and $K^{*'}$ vector resonances. As our central results, we quote the
values of table~\ref{tab4} at $s_{\rm cut}=4\;\gev^2$. To the uncertainty
given in table~\ref{tab4}, we add an error for the variation of our results
when changing $s_{\rm cut}$. The resonance mass and width parameters are then
found to be:
\begin{equation}
\label{mKsgaKs}
m_{K^*} \,=\, 943.41 \pm 0.59\;\mev \,, \qquad
\gamma_{K^*} \,=\, 66.72 \pm 0.87\;\mev \,,
\end{equation}
\begin{equation}
\label{mKspgaKsp}
m_{K^{*'}} \,=\, 1374 \pm 45\;\mev \,, \qquad
\gamma_{K^{*'}} \,=\, 240 \pm 131\;\mev \,,
\end{equation}
while the mixing parameter for the second resonance reads
$\gamma=-\,0.039\pm 0.020$. As has been already stated above, the quantities
of eqs.~\eqn{mKsgaKs} and \eqn{mKspgaKsp} are unphysical fit parameters.

To obtain physical parameters for the $K^*$ and $K^{*'}$ resonances, we have
to compute the positions of the poles of the vector form factor in the complex
$s$-plane. From the pole position $s_p$ we can then read off the physical mass
and width of the respective resonance according to the relation
\begin{equation}
\label{sp}
\sqrt{s_p} \,=\, M_R - \frac{i}{2}\,\Gamma_R \,.
\end{equation}
Calculating the pole positions along the lines of the approach outlined in
ref.~\cite{eglmp02} yields:
\begin{equation}
\label{MKsGaKs}
M_{K^*} \,=\, 892.01 \pm 0.92\;\mev \,, \qquad
\Gamma_{K^*} \,=\, 46.20 \pm 0.38\;\mev \,,
\end{equation}
\begin{equation}
\label{MKspGaKsp}
M_{K^{*'}} \,=\, 1276^{\,+\,72}_{\,-\,77}\;\mev \,, \qquad
\Gamma_{K^{*'}} \,=\, 198^{\,+\,61}_{\,-\,87}\;\mev \,.
\end{equation}
The uncertainties are calculated by assuming a Gaussian error propagation while
simultaneously varying both $m_R$ and $\gamma_R$. The mass of the charged $K^*$
meson turns out rather close to the value advocated by the PDG, but more than
$3\;\mev$ lower than the Breit-Wigner resonance parameters obtained in
refs.~\cite{belle07,jpp08}. On the other hand, the $K^*$ width $\Gamma_{K^*}$
of eq.~\eqn{MKsGaKs} nicely agrees with the result of \cite{belle07}, but it
is more than $4\;\mev$ lower than the PDG average.

To shed further light on the comparison with previous works let us calculate
the pole position of the $K^*(892)$ for the best fit of ref. \cite{jpp08}. The
model employed in this reference amounts, as far as the poles are concerned,
to removing the term proportional to $\RE\wt H_{K\pi}(s)$ from eq.~\eqn{Dden}
while keeping the energy dependent width. Denoting the respective fit
parameters with a tilde we have $\wt m_{K^*}= 895.28\pm 0.20\;\mev$ and
$\wt \gamma_{K^*}= 47.50\pm 0.41\;\mev$ \cite{jpp08}. The corresponding pole
position is given as the second line in table~\ref{tab5}, being perfectly
consistent with our results provided in eq.~\eqn{MKsGaKs} and the first line
of table~\ref{tab5}. The same exercise can be repeated for the original
analysis performed by the Belle collaboration \cite{belle07}. Here, we have
employed the fit parameters corresponding to the second fit given in table~3,
which are close to their final result for the $K^*$ parameters. In this case
the corresponding pole position is displayed as the last line in
table~\ref{tab5}. Again it turns out rather close to the previous results.
To summarise, the pole position is found to be rather stable since different
models yield compatible values for the {\it physical} parameters $M_{K^*}$ and
$\Gamma_{K^*}$ as defined in eq.~\eqn{sp}. Concerning the parameters of the
second resonance, they are in general agreement with the findings of
ref.~\cite{jpp08}, especially after the pole position is computed for the
latter results. Due to the large uncertainties, however, we cannot make any
more definite statement.

\begin{table}[htb]
\begin{center}
\vspace{3mm}
\begin{tabular}{c c c}
\hline \hline
&  Model Parameters &  Pole Positions \\
&     $(m_{K^*}, \gamma_{K^*})$ [MeV] & $(M_{K^*}, \Gamma_{K^*})$ [MeV] \\
\hline
This work           & $(943.41 \pm 0.59, 66.72 \pm 0.87)$ &
                      $(892.0  \pm 0.9,  46.2  \pm 0.4)$ \\
Ref.~\cite{jpp08}   & $(895.28 \pm 0.20, 47.50 \pm 0.41)$ &
                      $(892.1  \pm 0.2,  46.5  \pm 0.4)$ \\
Ref.~\cite{belle07} & $(895.47 \pm 0.20, 46.19 \pm 0.57)$ &
                      $(892.5  \pm 0.2,  45.3  \pm 0.5)$ \\
\hline \hline
\end{tabular}
\caption{Comparison between model parameters and corresponding pole positions
for the charged $K^*(892)$ meson. For definiteness, from ref.~\cite{belle07}
we have employed the parameters of the second fit of their table~3 and consider
only statistical uncertainties.\label{tab5}}
\end{center}
\end{table}

Our fits to the $\tau^-\to\nu_\tau K_S\pi^-$ spectrum also allow to determine
the slope parameters of the vector form factor $F_+^{K\pi}(s)$. The advantage
of using a three-subtracted dispersion relation is that the parameters
$\lambda_+^{'}$ and $\lambda_+^{''}$ are directly determined from the data,
making the extraction more model independent. The disadvantage being that
therefore the uncertainties for $\lambda_+^{'}$ turn out larger than for
example in ref.~\cite{jpp08}, where these parameters are a direct consequence
of the form factor model. Higher slope parameters can of course also be
calculated through dispersive integrals. For example in the case of
$\lambda_+^{'''}$ one has the relation:
\begin{equation}
\label{lappp}
\lambda_+^{'''} \,=\, \alpha_1^3 + 3\,\alpha_1\alpha_2 + M_{\pi^-}^6\,
\frac{6}{\pi} \!\int\limits^{s_{\rm cut}}_{s_{K\pi}} \!\!ds'\,
\frac{\delta_1^{K\pi}(s')}{(s')^4} \,.
\end{equation}
Together with the explicit fit results, this leads to
\begin{equation}
\label{lambdap}
\lambda_+^{'}   \,=\, (24.66 \pm 0.77)\cdot 10^{-3} \,, \;\;
\lambda_+^{''}  \,=\, (11.99 \pm 0.20)\cdot 10^{-4} \,, \;\;
\lambda_+^{'''} \,=\,  (8.73 \pm 0.16)\cdot 10^{-5} \,,
\end{equation}
where again the uncertainty due to the variation of $s_{\rm cut}$ has been
included in quadrature. Within the given errors, the value \eqn{lambdap} for
$\lambda_+^{'}$ is in good agreement to the result of ref.~\cite{jpp08}, as
well as the determination from an average of current experimental data for
$K_{l3}$ decays \cite{FKWG08}. On the other hand, both, the quadratic slope
$\lambda_+^{''}$, and the cubic slope $\lambda_+^{'''}$, are found somewhat
lower than the corresponding results of ref.~\cite{jpp08}.

To conclude, differential decay spectra of hadronic $\tau$ decays provide
important information for testing form factor models and extracting the
corresponding model parameters, thereby accessing QCD in the realm of low
energies. It will be very interesting to see if our findings are corroborated
by additional experimental data in the future. Furthermore, when comparing
parameters of hadronic resonances, even when employing Breit-Wigner type
parametrisations with an energy-dependent width, pole positions in the
complex $s$-plane should be provided in order to arrive at more model
independent results.

\vskip 1cm \noindent
{\Large\bf Acknowledgements}

\noindent
MJ is most grateful to the Belle collaboration, in particular S.~Eidelman,
D.~Epifanov and B.~Shwartz, for providing their data and for useful discussions.
He should also like to thank the referee of ref.~\cite{jpp08} for a question
which initiated the present study.
This work was supported in part by the Ramon y Cajal program (RE),
the Ministerio de Educaci\'on y Ciencia under grant FPA2005-02211,
the EU Contract No.~MRTN-CT-2006-035482, ``FLAVIAnet'',
the Spanish Consolider-Ingenio 2010 Programme CPAN (CSD2007-00042), and
the Generalitat de Catalunya under grant 2005-SGR-00994.


\begin{thebibliography}{10}

\bibitem{bnp92}
{\sc E.~Braaten}, {\sc S.~Narison}, and {\sc A.~Pich},
\newblock {QCD} analysis of the $\tau$ hadronic width,
\newblock {\em Nucl. Phys.} {\bf B373} (1992) 581.

\bibitem{bra89}
{\sc E.~Braaten},
\newblock The perturbative {QCD} corrections to the ratio ${R}$ for $\tau$
  decays,
\newblock {\em Phys. Rev.} {\bf D39} (1989) 1458.

\bibitem{bra88}
{\sc E.~Braaten},
\newblock {QCD} predictions for the decay of the $\tau$ lepton,
\newblock {\em Phys. Rev. Lett.} {\bf 60} (1988) 1606.

\bibitem{np88}
{\sc S.~Narison} and {\sc A.~Pich},
\newblock {QCD} formulation of the $\tau$ decay and determination of
  {${\Lambda}_{{\rm MS}}$},
\newblock {\em Phys. Lett.} {\bf B211} (1988) 183.

\bibitem{pich89}
{\sc A.~Pich},
\newblock QCD Tests from Tau-Decay Data,
\newblock Proc. Tau-Charm Factory Workshop (SLAC, Stanford, California, May
  23-27, 1989), SLAC report-343 (1989) 416.

\bibitem{ddhmz08}
{\sc M.~Davier}, {\sc S.~Descotes-Genon}, {\sc A.~H{\"o}cker}, {\sc
  B.~Malaescu}, and {\sc Z.~Zhang},
\newblock {The determination of $\alpha_s$ from $\tau$ decays revisited},
\newblock {\em Eur. Phys. J.} {\bf C56} (2008) 305,
\newblock arXiv:0803.0979 [hep-ph].

\bibitem{bj08}
{\sc M.~Beneke} and {\sc M.~Jamin},
\newblock {$\alpha_s$ and the $\tau$ hadronic width: fixed-order,
  contour-improved and higher-order perturbation theory},
\newblock {\em J. High Energy Phys.} {\bf 09} (2008) 044,
\newblock arXiv:0806.3156 [hep-ph].

\bibitem{mal08}
{\sc K.~Maltman} and {\sc T.~Yavin},
\newblock {$\alpha_s(M_Z)$ from hadronic $\tau$ decays},
\newblock (2008),
\newblock arXiv:0807.0650 [hep-ph].

\bibitem{aleph99}
{\sc R.~Barate} {\em et~al.},
\newblock Study of $\tau$ decays involving kaons, spectral functions and
  determination of the strange quark mass,
\newblock {\em Eur. Phys. J.} {\bf C11} (1999) 599,
\newblock hep-ex/9903015.

\bibitem{opal04}
{\sc G.~Abbiendi} {\em et~al.},
\newblock Measurement of the strange spectral function in hadronic $\tau$
  decays,
\newblock {\em Eur. Phys. J.} {\bf C35} (2004) 437,
\newblock hep-ex/0406007.

\bibitem{dhz05}
{\sc M.~Davier}, {\sc A.~H{\"o}cker}, and {\sc Z.~Zhang},
\newblock {The physics of hadronic $\tau$ decays},
\newblock {\em Rev. Mod. Phys.} {\bf 78} (2006) 1043,
\newblock hep-ph/0507078.

\bibitem{mwbnr08}
{\sc K.~Maltman}, {\sc C.~E. Wolfe}, {\sc S.~Banerjee}, {\sc I.~Nugent}, and
  {\sc J.~M. Roney},
\newblock {Status of the hadronic $\tau$ determination of $V_{us}$},
\newblock (2008),
\newblock arXiv:0807.3195.

\bibitem{gjpps07}
{\sc E.~G\'amiz}, {\sc M.~Jamin}, {\sc A.~Pich}, {\sc J.~Prades}, and {\sc
  F.~Schwab},
\newblock {Theoretical progress on the $V_{us}$ determination from $\tau$
  decays},
\newblock (2007),
\newblock {Proc. of {\em Kaon International Conference} (KAON'07), Frascati,
  Italy, 21-25 May 2007, arXiv:0709.0282 [hep-ph]}.

\bibitem{gjpps04}
{\sc E.~G\'amiz}, {\sc M.~Jamin}, {\sc A.~Pich}, {\sc J.~Prades}, and {\sc
  F.~Schwab},
\newblock ${V}_{us}$ and $m_s$ from hadronic $\tau$ decays,
\newblock {\em Phys. Rev. Lett.} {\bf 94} (2005) 011803,
\newblock Nucl. Phys. Proc. Suppl. {\bf 144} (2005) 59-64, hep-ph/0411278.

\bibitem{gjpps03}
{\sc E.~G\'amiz}, {\sc M.~Jamin}, {\sc A.~Pich}, {\sc J.~Prades}, and {\sc
  F.~Schwab},
\newblock Determination of $m_s$ and $|{V}_{us}|$ from hadronic $\tau$ decays,
\newblock {\em J. High Energy Phys.} {\bf 01} (2003) 060,
\newblock hep-ph/0212230.

\bibitem{pp98}
{\sc A.~Pich} and {\sc J.~Prades},
\newblock Perturbative quark mass corrections to the $\tau$ hadronic width,
\newblock {\em J. High Energy Phys.} {\bf 06} (1998) 013,
\newblock hep-ph/9804462, Nucl. Phys. Proc. Suppl. {\bf 74} (1999) 309, J.
  Prades, Nucl. Phys. Proc. Suppl. {\bf 76} (1999) 341.

\bibitem{ckp98}
{\sc K.~G. Chetyrkin}, {\sc J.~H. K{\"u}hn}, and {\sc A.~A. Pivovarov},
\newblock Determining the strange quark mass in {Cabibbo} suppressed $\tau$
  lepton decays,
\newblock {\em Nucl. Phys.} {\bf B533} (1998) 473,
\newblock hep-ph/9805335.

\bibitem{pp99}
{\sc A.~Pich} and {\sc J.~Prades},
\newblock Strange quark mass determination from {Cabibbo}-suppressed $\tau$
  decays,
\newblock {\em J. High Energy Phys.} {\bf 10} (1999) 004,
\newblock Nucl. Phys. Proc. Suppl. {\bf 86} (2000) 236, hep-ph/9909244.

\bibitem{kkp00}
{\sc J.~G. K{\"o}rner}, {\sc F.~Krajewski}, and {\sc A.~A. Pivovarov},
\newblock Determination of the strange quark mass from {Cabibbo} suppressed
  $\tau$ decays with resummed perturbation theory in an effective scheme,
\newblock {\em Eur. Phys. J.} {\bf C20} (2001) 259,
\newblock hep-ph/0003165.

\bibitem{km00}
{\sc J.~Kambor} and {\sc K.~Maltman},
\newblock The strange quark mass from flavor breaking in hadronic $\tau$
  decays,
\newblock {\em Phys. Rev.} {\bf D62} (2000) 093023,
\newblock hep-ph/0005156.

\bibitem{dchpp01}
{\sc M.~Davier}, {\sc S.~M. Chen}, {\sc A.~H{\"o}cker}, {\sc J.~Prades}, and
  {\sc A.~Pich},
\newblock Strange quark mass from $\tau$ decays,
\newblock {\em Nucl. Phys. Proc. Suppl.} {\bf 98} (2001) 319.

\bibitem{cdghpp01}
{\sc S.~M. Chen}, {\sc M.~Davier}, {\sc E.~G\'amiz}, {\sc A.~H{\"o}cker}, {\sc
  A.~Pich}, {\em et~al.},
\newblock Strange quark mass from the invariant mass distribution of
  {Cabibbo}-suppressed $\tau$ decays,
\newblock {\em Eur. Phys. J.} {\bf C22} (2001) 31,
\newblock hep-ph/0105253.

\bibitem{bck04}
{\sc P.~A. Baikov}, {\sc K.~G. Chetyrkin}, and {\sc J.~H. K{\"u}hn},
\newblock Strange quark mass from $\tau$ lepton decays with {${\cal
  O}(\alpha_s^3)$} accuracy,
\newblock {\em Phys. Rev. Lett.} {\bf 95} (2005) 012003,
\newblock hep-ph/0412350.

\bibitem{belle07}
{\sc D.~Epifanov} {\em et~al.},
\newblock Study of $\tau^- \to K_S\pi^-\nu_\tau$ decay at Belle,
\newblock {\em Phys. Lett.} {\bf B654} (2007) 65,
\newblock arXiv:0706.2231 [hep-ex].

\bibitem{babar07}
{\sc B.~Aubert} {\em et~al.},
\newblock {Measurement of the $\tau^-\to K^-\pi^0\nu_\tau$ Branching Fraction},
\newblock {\em Phys. Rev.} {\bf D76} (2007) 051104,
\newblock arXiv:0707.2922 [hep-ex].

\bibitem{babar08}
{\sc B.~Aubert} {\em et~al.},
\newblock {Measurement of the $B(\tau^-\to\bar K^0\pi^-\nu_\tau)$ using the
  BaBar detector},
\newblock (2008),
\newblock talk presented at ICHEP08, Philadelphia, Pensylvania,
  arXiv:0808.1121v2 [hep-ex].


\bibitem{wren08}
{\sc A.~Wren},
\newblock {Precision measurement of the branching fraction of $\tau^-\to\bar
  K^0\pi^-\nu_\tau$ at BaBar},
\newblock (2008),
\newblock talk presented at Tau08, Novosibirsk, Russia.



\bibitem{fm96}
{\sc M.~Finkemeier} and {\sc E.~Mirkes},
\newblock The scalar contribution to $\tau\to K\pi\nu_\tau$,
\newblock {\em Z. Phys.} {\bf C72} (1996) 619,
\newblock hep-ph/9601275.

\bibitem{erl02}
{\sc J.~Erler},
\newblock {Electroweak radiative corrections to semileptonic tau decays},
\newblock {\em Rev. Mex. Fis.} {\bf 50} (2004) 200,
\newblock hep-ph/0211345.

\bibitem{egpr89}
{\sc G.~Ecker}, {\sc J.~Gasser}, {\sc A.~Pich}, and {\sc E.~{de Rafael}},
\newblock The role of resonances in chiral perturbation theory,
\newblock {\em Nucl. Phys.} {\bf B321} (1989) 311.

\bibitem{eglpr89}
{\sc G.~Ecker}, {\sc J.~Gasser}, {\sc H.~Leutwyler}, {\sc A.~Pich}, and {\sc
  E.~{de Rafael}},
\newblock Chiral Lagrangians for massive spin 1 fields,
\newblock {\em Phys. Lett.} {\bf B223} (1989) 425.

\bibitem{jpp06}
{\sc M.~Jamin}, {\sc A.~Pich}, and {\sc J.~Portol\'es},
\newblock Spectral distribution for the decay $\tau\to\nu_\tau K\pi$,
\newblock {\em Phys. Lett.} {\bf B~640} (2006) 176,
\newblock hep-ph/0605096.

\bibitem{gp97}
{\sc F.~Guerrero} and {\sc A.~Pich},
\newblock Effective field theory description of the pion form factor,
\newblock {\em Phys. Lett.} {\bf B412} (1997) 382,
\newblock hep-ph/9707347.

\bibitem{pp01}
{\sc A.~Pich} and {\sc J.~Portol\'es},
\newblock The vector form factor of the pion from unitarity and analyticity: A
  model-independent approach,
\newblock {\em Phys. Rev.} {\bf D63} (2001) 093005,
\newblock hep-ph/0101194.

\bibitem{scp02}
{\sc J.~J. Sanz-Cillero} and {\sc A.~Pich},
\newblock Rho meson properties in the chiral theory framework,
\newblock {\em Eur. Phys. J.} {\bf C27} (2003) 587,
\newblock hep-ph/0208199.

\bibitem{jpp08}
{\sc M.~Jamin}, {\sc A.~Pich}, and {\sc J.~Portol\'es},
\newblock What can be learned from the Belle spectrum for the decay
  $\tau\to\nu_\tau K_S\pi^-$,
\newblock {\em Phys. Lett.} {\bf B664} (2008) 78,
\newblock arXiv:0803.1786 [hep-ph].

\bibitem{jop00}
{\sc M.~Jamin}, {\sc J.~A. Oller}, and {\sc A.~Pich},
\newblock S-wave $K\pi$ scattering in chiral perturbation theory with
  resonances,
\newblock {\em Nucl. Phys.} {\bf B587} (2000) 331,
\newblock hep-ph/0006045.

\bibitem{jop01a}
{\sc M.~Jamin}, {\sc J.~A. Oller}, and {\sc A.~Pich},
\newblock Strangeness-changing scalar form factors,
\newblock {\em Nucl. Phys.} {\bf B622} (2002) 279,
\newblock hep-ph/0110193.

\bibitem{jop01b}
{\sc M.~Jamin}, {\sc J.~A. Oller}, and {\sc A.~Pich},
\newblock Light quark masses from scalar sum rules,
\newblock {\em Eur. Phys. J.} {\bf C24} (2002) 237,
\newblock hep-ph/0110194.

\bibitem{jop06}
{\sc M.~Jamin}, {\sc J.~A. Oller}, and {\sc A.~Pich},
\newblock Scalar $K\pi$ form factor and light quark masses,
\newblock {\em Phys. Rev.} {\bf D~74} (2006) 074009,
\newblock hep-ph/0605095.

\bibitem{mou07}
{\sc B.~Moussallam},
\newblock {Analyticity constraints on the strangeness changing vector current
  and applications to $\tau\to K\pi \nu_\tau$, $\tau\to K\pi\pi \nu_\tau$},
\newblock {\em Eur. Phys. J.} {\bf C53} (2008) 401,
\newblock arXiv:0710.0548 [hep-ph].

\bibitem{om58}
{\sc R.~Omn\`es},
\newblock On the Solution of certain singular integral equations of quantum
  field theory,
\newblock {\em Nuovo Cim.} {\bf 8} (1958) 316.

\bibitem{pp00}
{\sc E.~Pallante} and {\sc A.~Pich},
\newblock {Final state interactions in kaon decays},
\newblock {\em Nucl. Phys.} {\bf B592} (2001) 294,
\newblock hep-ph/0007208.

\bibitem{oor00}
{\sc J.~A. Oller}, {\sc E.~Oset}, and {\sc A.~Ramos},
\newblock {Chiral unitary approach to meson meson and meson baryon interactions
  and nuclear applications},
\newblock {\em Prog. Part. Nucl. Phys.} {\bf 45} (2000) 157,
\newblock hep-ph/0002193.

\bibitem{Gounaris:1968mw}
{\sc G.~J.~Gounaris} and {\sc J.~J.~Sakurai},
\newblock {Finite width corrections to the vector meson dominance prediction for rho $\to$ e+ e-}, \newblock {\em Phys. Rev. Lett.} {\bf 21} (1968) 244.
  
\bibitem{Kuhn:1990ad}
{\sc J.~H.~K{\"u}hn} and {\sc A.~Santamaria},
\newblock {Tau decays to pions},
\newblock {Z. Phys.} {\bf C48} (1990) 445.
  
\bibitem{FKWG08}
{\sc M.~Antonelli} {\em et~al.},
\newblock {Precision tests of the Standard Model with leptonic and semileptonic
  kaon decays},
\newblock (2008),
\newblock arXiv:0801.1817 [hep-ph].

\bibitem{epi07}
{\sc D.~Epifanov},
\newblock email communication.

\bibitem{eglmp02}
{\sc R.~Escribano}, {\sc A.~Gallegos}, {\sc J.~L. Lucio~M}, {\sc G.~Moreno},
  and {\sc J.~Pestieau},
\newblock {On the mass, width and coupling constants of the $f_0(980)$},
\newblock {\em Eur. Phys. J.} {\bf C28} (2003) 107,
\newblock hep-ph/0204338.

\end{thebibliography}

\end{document}